# Ferromagnetism in lead graphite-pencils and magnetic composite with $CoFe_2O_4$ particles

R.N. Bhowmik

Department of Physics, Pondicherry University, R. Venkataraman Nagar, Kalapet, Pondicherry-60014, India.

Correspondence: Tel.:+91-9944064547; fax: +91-4132655734.

E-mail: rnbhowmik.phy@pondiuni.edu.in

**Abstract**

This work has been initiated with a curiosity to investigate the elemental composition and magnetic response of different grades of lead pencils (6B, 2B, HB, 2H, 5H) that people use in daily life. Interestingly, experimental results landed with a great achievement of observing soft magnetism in lead pencils, indicating a wide scope of magnetic tuning for room temperature applications. A novel magnetic composite has been synthesized by mixing different concentration of $CoFe_2O_4$ (CF) nanoparticles in 5H and 6B pencils for studying the magnetic tailoring aspects using pencils. Our results showed different possibilities of controlling disorder induced ferromagnetic parameters and a simple approach of producing sufficiently high coercive magnetic composite using pencils.

**Key words:** Carbon-magnetic composite, Soft ferromagnet, Magnetic tailoring, Magnetic grains, room temperature ferromagnetism.

## 1. Introduction

Recent trend of magnetic materials research is focusing on two basic aspects. First one is the fundamental study and development of novel materials with elemental and chemical homogeneity down to atomic scale. The synthesis of true homogeneous material in ambient conditions is not only difficult, but also not suitable in most of the cases for practical fields. Secondly, hetero-structured materials or composites, consisting of the dispersion of magnetic particles in organic and inorganic matrix, has received tremendous research interest due to the ability of tailoring physical properties in designing multifunctional materials [1]. In order to design ferromagnetic materials for room temperature spintronic applications, different materials, which are heterogeneous in true sense, have been suggested in literature. Dilute magnetic semiconductor [2, 3] and carbon (graphite) [4, 5] based materials are the most exciting in the sense that they exhibited ferromagnetism at room temperature and above, which is unconventional and needs details understanding. The basic fact is that origin of the ferromagnetism in both type of materials were realized to be affected by coexisting disorder in the lattice structure [6, 7].

Lead pencil is one of the easily available hetero-structured graphite based materials that people use in their daily life for writing, sketching, coloring, designing, makeup, technical field, etc. Lead pencil, as the name suggests, does not contain metallic lead in it. Rather Lead pencil is an example of intercalated compound, consisting of the mixture of clay (mainly $SiO_2$ and minor amount of metal oxides) particles in conducting graphite matrix. Lead pencils are sub-classified depending on their darkness (quantity of graphite) and hardness (quantity of impurity) and a scale can be formed ranging from 9H to 9B, where H represents hardness and B represents blackness. As we follow the notation 6B, HB and 5H, the content of clay in the pencil increases. In this work we treat clay particles as the guest, whose contribution is to increase the disorder in the host of layered structure of graphite [8]. During the accommodation of guest atoms one can expect carbon atom vacancy and lattice defects in the in-plane and out of plane structure of graphite. If so, some uncompensated electronic spins are also expected that can exhibit ferromagnetism in the graphite structure [9].

Realization of the mechanism of ferromagnetism in graphite structure is a challenging task to researchers, because it contains only s and p electrons in contrast to 3d and 4f electrons in traditional ferromagnets. Knowledge of the graphite based ferromagnets will show huge impact towards the development of graphite based materials for spintronics, sensors, biomedicines, telecommunications and automobile sector [5, 10, 11]. The bonding between carbon atoms in graphite layers involves $sp^2$ hybridization and formation of in-plane $\sigma$ band. The unaffected p orbital of each carbon atom, which is perpendicular to the plane, forms a $\pi$ band that binds covalently the carbon atoms of neighboring layers. Half filled $\pi$ band and the weak van der Waals coupling between the carbon layers in graphite permit the capture of foreign atoms and molecular species in between the layers to form the graphite intercalated compound [12]. We believe the observed ferromagnetism in graphite is most probably affected by the out of plane electronic spin structure of graphite [13], like the origin of ferromagnetism above Morin transition temperature (~ 270 K) in $\alpha$-$Fe_2O_3$ [14]. The distance between two layers of the graphite structure may be changed with the guest size and provide a good model systems for studying magnetism and electrical transport in the presence of different guests.

The present work was motivated from the fact that disorder and lattice defects at the edge and grain boundaries of graphite structure can produce ferromagnetism [4,5]. Indeed, we have room temperature ferromagnetism in lead pencils. Our aim is to focus on the ferromagnetic properties as a function of increasing disorder due to increasing clay content in graphite matrix. Next, we want to examine the tailoring of ferromagnetism in lead pencil matrix by mixing different quantity of $CoFe_2O_4$ nanoparticles. $CoFe_2O_4$ is selected for its intrinsic nature of exhibiting magnetic hardness with high cubic magneto crystalline anisotropy, high coercivity, and moderate saturation magnetization.

## 2. Experimental

We have procured from market the lead pencils of model Apsara 6B, 2B, HB, 2H and 5H made by Hindustan Pencil Ltd., Mumbai, India. The graphite core of the pencil was carefully taken out and made into powders using agate mortar and pestle. A portion of the ground powder of each model was kept for experimental studies. Since $SiO_2$ is the major constituent of clay in lead pencil, we have prepared two extreme end models 6BX and 5HX by mixing standard graphite powder (from Sigma Aldrich) and $SiO_2$ (purity 99.9%) for studying the non-magnetic $SiO_2$ effect in graphite matrix. We mixed graphite and $SiO_2$ masses with weight ratio 84:16 and 52: 48 to get 6BX and 5HX, respectively. In this case we have followed general weight ratio of graphite and clay in 6B and 5H lead pencils. In order to study the coexisting magnetic particles in lead pencil matrix, we have mixed the ground powder of 6B and 5H separately with magnetic material $CoFe_2O_4$ (CF). In this magnetic composite the weight ratio of lead pencil and CF were maintained to 75:25, 50:50 and 25:75, respectively. The magnetic composites were denoted as 6BCF25, 6BCF50, 6BCF75 and 5HCF25, 5HCF50, 5HCF75 for 6B and 5H series, respectively.

The mixture was made into pellet after proper grinding. It may be informed that $CoFe_2O_4$ (CF) with grain size ~10 nm was separately prepared by 100 hours mechanical milling of micron grain sized $CoFe_2O_4$ using Pulverisette mono miller (FRITSCH P6, Germany).

Crystalline structure of the samples was examined from room temperature XRD spectrum using Xpert Panalytical X-ray diffractometer with $CuK_\alpha$ radiation ($\lambda$=1.54056 Å) in the 2θ range 10 to $90^0$ with step size $0.02^0$ and time per step 2 seconds. Magnetic field dependent magnetization of the samples was studied at room temperature using VSM (Lakeshore 7400). Temperature dependence of magnetization in zero field cooled (ZFC) and field cooled (FC) modes has been measured using MPMS7 SQUID magnetometer (Quantum Design, USA). Elemental concentration of the lead pencils was measured using Bruker S4 pioneer wavelength dispersive X-ray fluorescence spectrometer (WDXRF). Surface morphology of the samples was studied using Scanning Electron Microscope (HITACHI S-3400N). Energy dispersive X-ray (EDX) technique was used to cross check the elemental composition of the material using different micro analysis options, e.g., area analysis and point and shoot analysis (8-10 points).

**3. Results and discussion**

3.1. *Structural study*

Fig. 1 (a) shows the XRD spectra of selected lead pencils. The major component of the XRD spectra of pencils (e.g., 6B, HB, 5H) matched to hexagonal crystal structure with space group p63mmc of standard graphite structure. The impurity phase of clay in graphite matrix of the lead pencil was within the back ground when intensity of the spectrum was plotted in linear scale. Hence, XRD intensity in Fig. 1(a) is shown in log-scale visualize the XRD pattern of clay phase. We have seen $SiO_2$ as the major phase of clay in lead pencils (standard pattern of SiO$_2$ with reference code: 00-002-0471 has been shown in Fig. 1a). XRD spectra were analyzed using the peaks of graphite alone and the structural parameters are given in Table 1. We have found that lattice parameters (a = 2.4656 Å, c = 6.7378 Å, V = 35.47 Å$^3$) of pure graphite is close to the standard pattern (a = 2.4700 Å, c = 6.8000 Å, reference code: 00-001-0640 has been shown in Fig. 1a). We have noted that lattice parameters in lead pencils, especially c axis is shrinking with the increase of clay content in the graphite matrix of lead pencil. This is the effect of intercalation of clay in between graphite planes of the hexagonal structure [15]. The intensity of the strongest (002) XRD peak of graphite phase decreases with the increase of clay content in lead pencil and Intensity counts are given in Table 1 for selected samples.

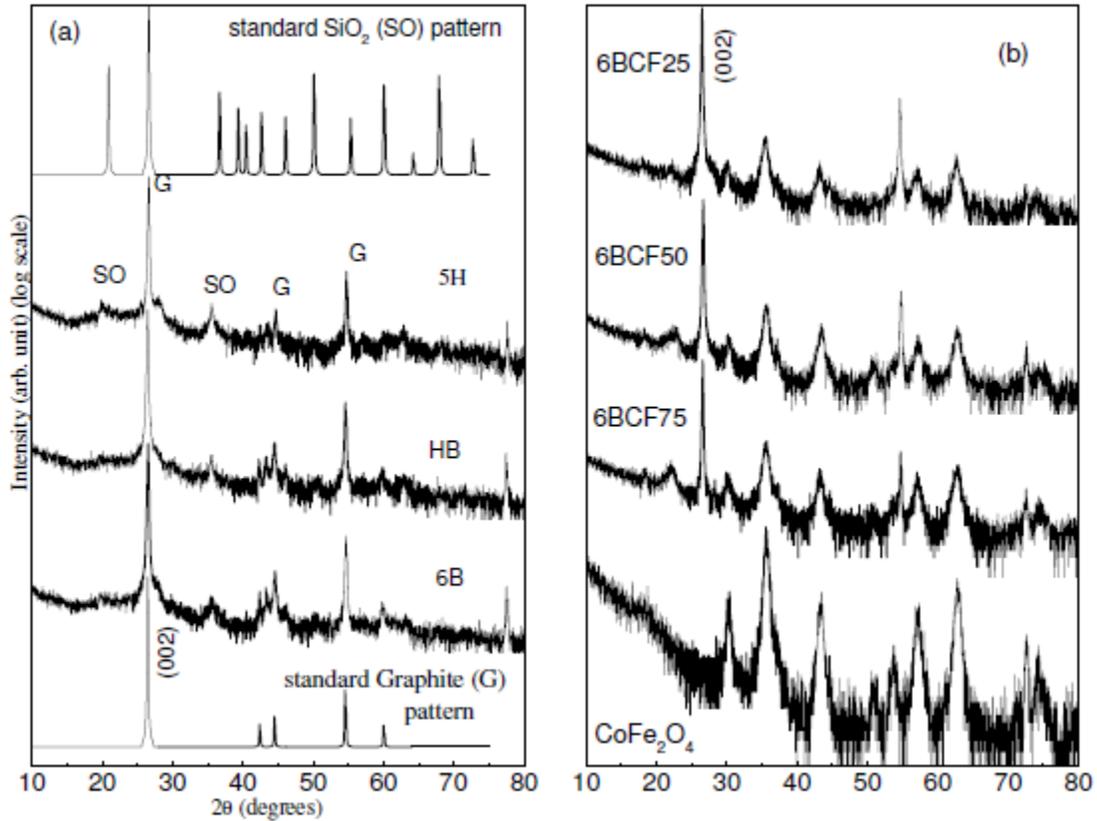

Fig.1 XRD spectra of pencils is compared with graphite and SiO$_2$ (in a) and CoFe$_2$O$_4$ mixed 6B samples (in b).

Table 1. Structural parameters of selected samples.

| Sample Name | a=b (Å) | c (Å) | V (Å$^3$) | (002) graphite peak counts | Composite sample | (002) graphite peak counts |
|---|---|---|---|---|---|---|
| 6B | 2.4646 | 6.7224 | 35.36 | 14710 | 6BCF25 | 7680 |
| HB | 2.4640 | 6.7221 | 35.34 | 13080 | 6BCF50 | 3900 |
| 5H | 2.4630 | 6.7193 | 35.30 | 10190 | 6BCF75 | 2020 |

Fig. 1(b) shows the XRD spectra of 6B composite (6BCF25, 6BCF50, 6BCF75) samples. In magnetic composites, XRD peak intensity of CF increases with the increase of $CoFe_2O_4$ content, irrespective of mixing with 6B and 5H lead pencils. It may be noted that XRD spectrum of $CoFe_2O_4$ ferrite was fitted with cubic spinel phase with space group Fd3m and lattice parameter (a =8.3615 Å). The increase of CF content in the lead pencil is confirmed from the decrease of XRD peak intensity of graphite phase, as shown for 6B mixed samples in Table 1. Elemental concentration of different lead pencils was estimated from XRF spectra. The estimated elemental composition of different samples is shown in Table 2. XRF data identified C, Si and O as the major elements of lead pencils. The basic information is that graphite (C) concentration decreases, where as the Silicon (Si) and Oxygen (O) increases with the increase of Clay content in graphite. A significant amount of Fe (Iron) and Al (Aluminium) is also noted and their concentration increases with the increase of clay content. Some minor elements, e.g., Na, Ca, Mg, K, S and Ti, were also noted as the constituent of clay below 0.01 concentrations. On the other hand, a spatial distribution of the elements was confirmed from the EDX spectra. The average values of the elements over 10 points of the scanned area are shown in Table 3.

Table 2. Elemental concentration % in different lead pencils from XRF data.

| Sample | C (±3.5) | Si (±0.2) | O (±1.0) | Fe (±0.1) | Al (±0.3) | Ca (±0.6) | K (±1.0) | Mg (±0.5) | Na (±1.0) | S (±0.5) | Ti (±0.8) |
|---|---|---|---|---|---|---|---|---|---|---|---|
| 6B | 60.3 | 11.1 | 18.0 | 3.6 | 3.7 | 0.9 | 0.4 | 0.7 | 0.5 | 0.3 | 0.3 |
| 2B | 51.4 | 13.9 | 20.3 | 5.5 | 5.0 | 1.0 | 0.4 | 0.9 | 0.6 | 0.2 | 0.4 |
| HB | 43.5 | 17.0 | 23.1 | 6.4 | 5.8 | 1.4 | 0.2 | 1.0 | 0.6 | 0.1 | 0.5 |
| 2H | 35.1 | 20.0 | 26.7 | 6.9 | 6.8 | 1.4 | 0.2 | 1.1 | 0.8 | 0.1 | 0.5 |
| 5H | 27.4 | 22.6 | 30.4 | 7.2 | 7.3 | 1.0 | 0.4 | 1.4 | 0.9 | 0.2 | 0.6 |

Table 3. Elemental composition in weight % of selected lead pencils from EDX data.

| Sample | C | Si | O | Fe | Al | Ca | Mg | Ti |
|---|---|---|---|---|---|---|---|---|
| 6B | 82.9 | 6.5 | 6.0 | 2.2 | 2.4 | -- | -- | -- |
| HB | 57.4 | 11.4 | 23.1 | 5.5 | 3.1 | 0.2 | 1.1 | 0.5 |
| 5H | 45.3 | 15.5 | 26.8 | 6.3 | 4.1 | -- | 1.3 | 0.5 |

In addition to the major elements (C, Si, O), significant amount of Fe and Al were also detected from EDX spectra. The basic trend of varying the elemental composition of graphite (Carbon) and clay in lead pencils is identical in both XRD and EDX data. However, semi-quantitative results of the major elements from XRF data, in particular Carbon (C), are not well consistent to the standard specific values, e.g., 84%, 68% and 52% C for pencils 6B, HB and 5H, respectively. This means XRF technique is good in identifying the trace elements which can not be determined by EDX. On the other hand, EDX results are close to the expected values of the major elements in different lead pencils and can be used for quantitative estimation of different elements in lead pencils. One typical EDX spectrum of 6B sample, showing major elements, is presented in Fig. 2. Inset of Fig. 2 shows the incorporating magnetic (CF) particles in the 6B pencil.

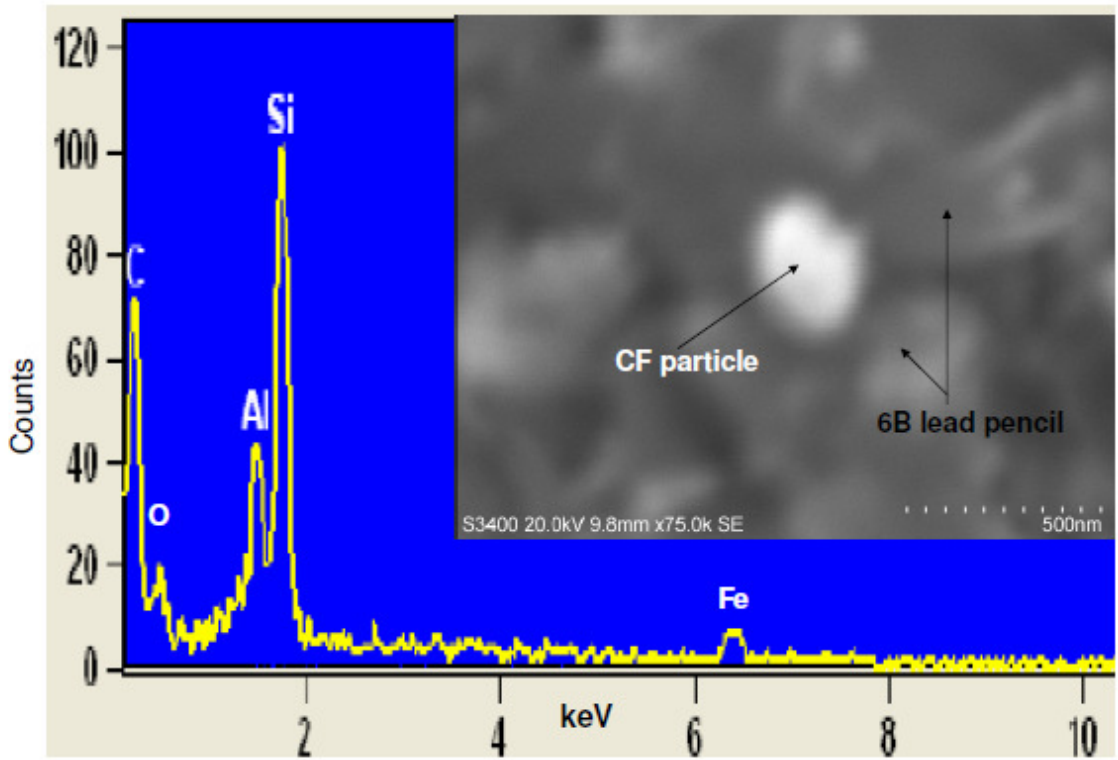

Fig. 2 (Colour online) EDX spectrum of 6B sample and SEM picture of 6BCF25 sample (inset).

**3.2.** *Magnetic properties*

Room temperature magnetic properties of different samples were examined from the field dependence of dc magnetization. Before magnetic description of the lead pencil and CF mixed composites, we have checked magnetic response of graphite, $SiO_2$ and

graphite-$SiO_2$ mixtures. Graphite is basically a diamagnet at room temperature (main panel of Fig. 3 (a)), but ferromagnetic signature (loop) is clearly coexisting in the low field range. Our observation is consistent to the reported feature of ferromagnetism in graphite [15, 16]. Magnitude of the remanent magnetization (~ 2.16 memu/g and 4.53 memu/g for samples 6BX and 5HX, respectively) in our samples are relatively small in comparison with ~ 40 memu/g that reported in bulk ferromagnetic graphite [15], but well comparable to some of the reported graphite samples (~3.2 memu/g) [16]. However, coercivity (~350 Oe and 280 Oe calculated from magnetization curve intercept in negative field axis for samples 6BX and 5HX, respectively) values are comparable to the value 350 Oe that was reported in bulk ferromagnetic graphite [15]. Our study shows that ferromagnetic component increases in unusual manner by increasing the content of $SiO_2$ in graphite, although $SiO_2$ has shown a typical diamagnetic feature (Inset of Fig. 3 (a)). As seen in Fig. 3 (b), ferromagnetic loop of 5HX sample (graphite 52 % and $SiO_2$ 48 %) is significantly large in comparison with 6BX sample (graphite 84 % and $SiO_2$ 16 %). This means increasing disorder in the graphite structure in the presence of $SiO_2$ definitely played important role in enhancing the ferromagnetic properties of graphite. Based on this result, we believe that unconventional ferromagnetism in graphite and some other materials is perhaps the extrinsic property and affected mostly at the grain boundaries and interfacial disordered structures. Hereafter, we understand the effect of disorder and structural inhomogeneity on the ferromagnetism of graphite using different lead pencils. Fig. 4 shows a systematic and dramatic enhancement of ferromagnetism when the clay content increases in lead pencils (6B → 2B → HB → 2H → 5H). Ferromagnetic loop is nearly symmetric for all lead pencils, as shown in the inset of Fig. 4. The pencils are typical soft ferromagnets, showing rapid increase of magnetization at the initial stage of magnetic field increase. This is followed by magnetic saturation above 2 kOe. As an effect of increasing clay content in the graphite matrix, we noted the following features: (1) small diamagnetic contribution (i.e., decrease of magnetization with the increase of field above 5 kOe) in 6B and 2B samples, (2) magnetic non-saturation (i.e., non-linear increase of magnetization with field above 2 kOe) in HB, 2H, 5H samples, (3) increase of ferromagnetic loop. The magnetic parameters, e.g., spontaneous magnetization ($M_S$: calculated from the extrapolation of high field M(H) data to zero field limit) and remanent magnetization ($M_R$: magnetization retained in the sample after reducing the field from 15 kOe to 0 Oe) have enhanced remarkably with the increase of clay content. It is true that magnetic component due to presence of Fe was noted in clay and the Fe concentration gradually increases while the pencil changes from B series to H series. The observation of enhanced ferromagnetism in graphite due to increasing fraction of $SiO_2$ (Fig. 3(b)) suggests that ferromagnetism in lead pencils is not solely due to the presence of magnetic impurity Fe. Rather, ferromagnetism in lead pencils is largely controlled by increasing disorder at the grain boundary spin structure of graphite particles. The magnetic impurity Fe is acting as the catalyst for the further enhancement of soft ferromagnetism with increasing moment (~ 1 emu/g for 5H sample), which is remarkably high in comparison with bulk magnetic graphite (~ 0.3 emu/g) [17]. Important magnetic parameters of different samples are shown in Table 4. It is interesting to note that change of coercivity ($H_C$: negative magnetic field needed to reduce the remanent magnetization zero) of the pencils with the increase of clay is not monotonic. Rather it decreases in B series, showing minimum value of $H_C$ for HB sample and $H_C$ once again increases on further increase of the clay in the H series. A comparative plot of the initial magnetization curves (Fig. 5) shows that field dependent increase of magnetization of the pencils are controlled mainly by magnetic domain rotation, where as grain boundary disorder and domain wall contribution slowly increases in the field dependent magnetization of $CoFe_2O_4$ (CF) nanoparticles and magnetic composites (CF mixed 6B and 5H samples). It is true that none of the composite samples have achieved the magnetization of CF sample. As shown in Fig. 6, the ferromagnetic development in 6B (Fig. 6(a)) and 5H (Fig. 6(b)) composites are drastically different from the lead pencil. One can tune magnetization (spontaneous magnetization and remanent magnetization) in the composite samples by increasing the content of CF particles. It is noted that $M_R/M_S$ ratio of the pencils are in the range 0.22-0.25 and 0.33 for the CF nanoparticle sample. This ratio can be enhanced in the range 0.37-0.39 by suitable mixing of lead pencil in CF matrix.

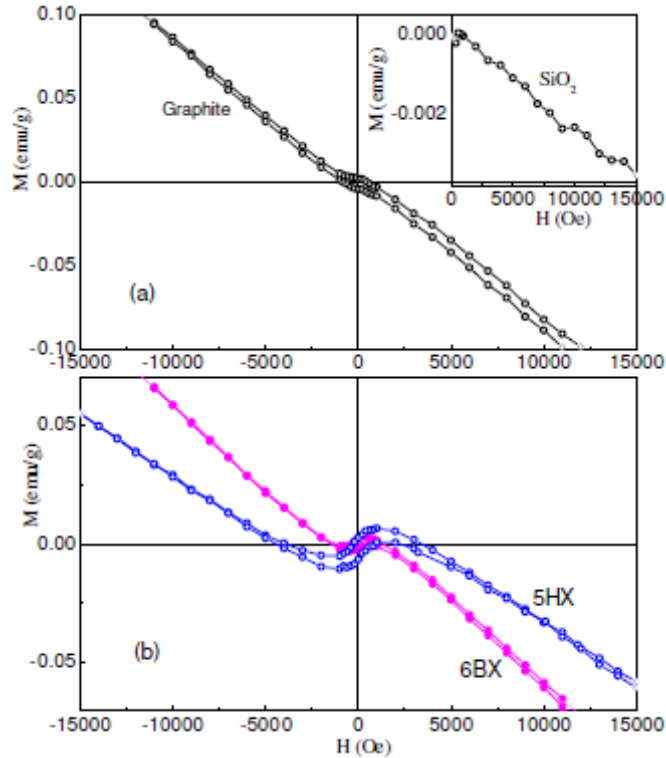

Fig. 3 (Colour online) Field dependent magnetization of standard Graphite (main frame of (a)) and $SiO_2$ (inset of (a)), and graphite-$SiO_2$ samples (in (b).

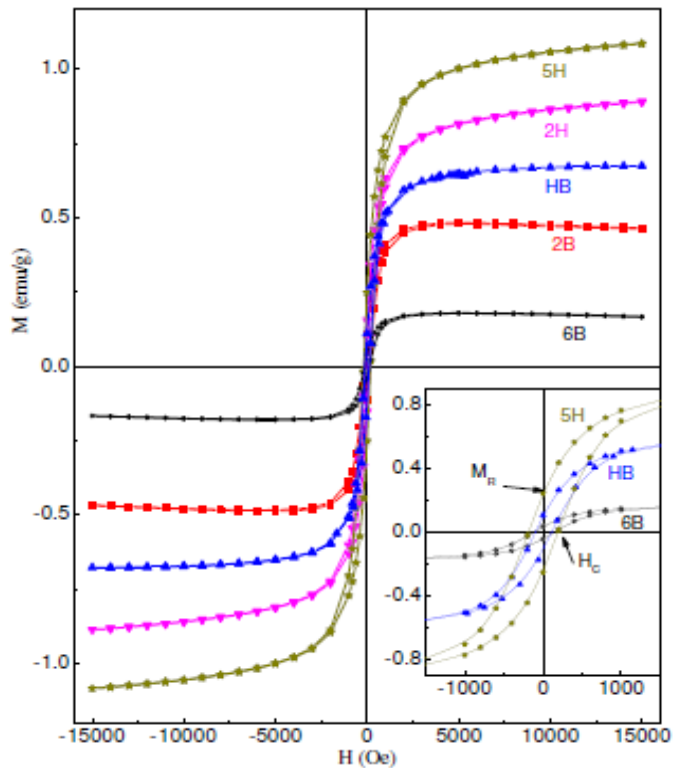

Fig. 4 (Colour online) Magnetic field dependence of magnetization of different lead pencils. Inset shows the hysteresis loop of selected samples.

Table 4. Different magnetic parameters from room temperature M(H) data

| Sample | $M_S$ (emu/g) | $M_R$ (emu/g) | $M_R/M_S$ | $H_C$ (Oe) |
|---|---|---|---|---|
| 6B | 0.175 | 0.038 | 0.22 | 130 |
| 2B | 0.480 | 0.110 | 0.23 | 125 |
| HB | 0.650 | 0.112 | 0.17 | 115 |
| 2H | 0.800 | 0.155 | 0.19 | 122 |
| 5H | 1.000 | 0.252 | 0.25 | 186 |
| $CoFe_2O_4$ | 49.260 | 16.540 | 0.33 | 890 |
| 6BCF25 | 9.350 | 3.650 | 0.39 | 965 |
| 6BCF50 | 16.184 | 5.430 | 0.33 | 800 |
| 6BCF75 | 28.610 | 10.440 | 0.36 | 930 |
| 5HCF25 | 13.420 | 4.660 | 0.35 | 680 |
| 5HCF50 | 20.000 | 6.840 | 0.34 | 760 |
| 5HCF75 | 33.836 | 12.560 | 0.37 | 915 |

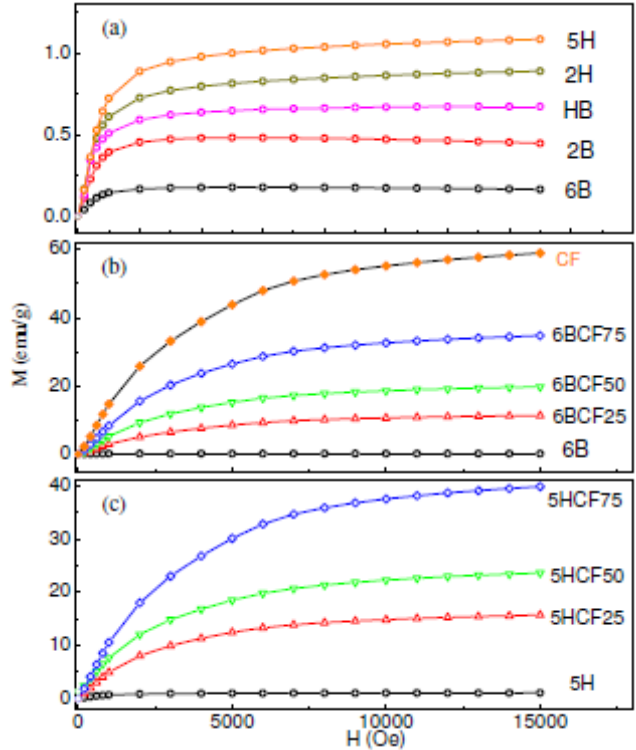

Fig.5 (Colour online) Magnetic field dependent magnetization of different lead pencils (in a), CF mixed 6B samples (in b) and CF mixed 5H samples (in c).

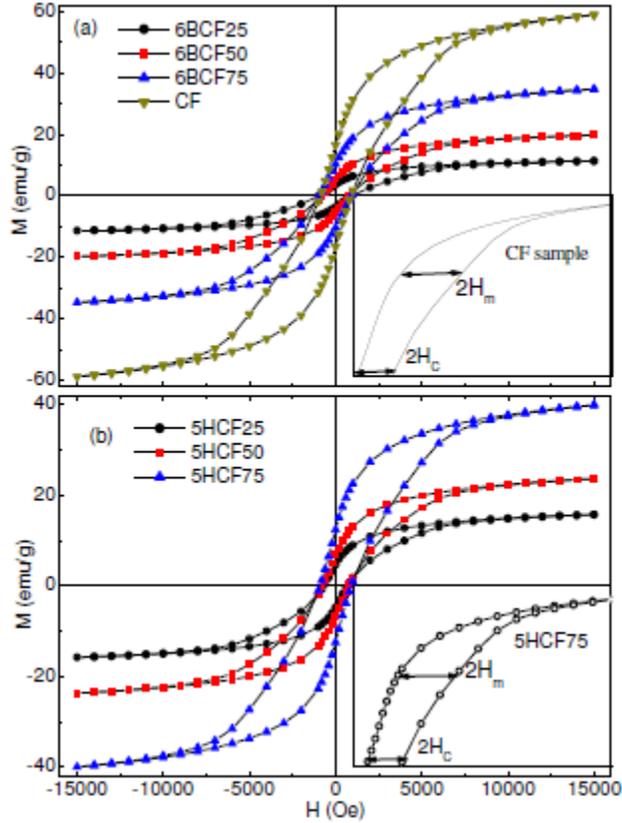

Fig. 6 (Colour online) M(H) loops of CF mixed 6B (in a) and 5H (in b) samples. Inset figures show the different loop widths.

Magnetic composites of $CoFe_2O_4$ particles also showed improvement of $M_R/M_S$ ratio in a wide range 0.2 to 0.65 depending on the synthesis condition and nature of composite matrix [18, 19, 20]. Room temperature coercivity ($H_C$ ~ 890 Oe) and spontaneous magnetization ($M_S$ ~50 emu/g) of used CF sample is within the range of reported value of $H_C$ ~ 750 Oe- 1000 Oe and $M_S$ ~50 to 60 emu/g for $CoFe_2O_4$ particles [18, 21, 22]. Although spontaneous magnetization of our composite samples are less ($M_S$ ~10 to 34 emu/g) than the CF sample due to large fraction of coexisting lead pencils, but one could see that $H_C$ of 6BCF25 composite sample is the largest (~ 965 Oe) among all samples. $H_C$

value then decreased for 6BCF50 sample below CF level. $H_C$ value for 6BCF75 sample once again increases above CF level, but this value is lower than 6BCF25 sample. 5H composites are interesting in the sense that both $M_S$ and $H_C$ of the mixed samples will be controlled monotonically over a wide range ($M_S$: 13.42 emu/g- 33.84 emu/g; $H_C$: 680 Oe -915 Oe) by increasing the CF content. Besides the magnetic parameters tuning, from physics point of view also we have noted two major changes in the magnetic properties of pencils after mixing with $CoFe_2O_4$ nanoparticles. First one is the loss of soft ferromagnetism observed for pencils and appearance of coexisting soft and hard magnetic phases. Nano-sized $CoFe_2O_4$ grains (inset of Fig. 6(a)) showed this typical feature, which is dominating in the mixed samples also. Inset of Fig. 6(b) gives an example for 5HCF75 sample. The coexistence of soft and hard magnetic phases has been observed in many magnetic composites due to core-shell structure [23]. The coexistence of soft and hard magnetic phases is understood from the differences between loop width on the field axis (i.e, $2H_C$ ~ 1.78 kOe and 1.83 kOe for CF and 5HCF75 samples, respectively) and distorted loop width (i.e., $2H_m$ ~ 3.13 kOe and 3.20 kOe for CF and 5HCF75 samples, respectively) at the magnetization nearly 40% below of the magnetization at 15 kOe. The whole process of tailoring the room temperature ferromagnetism is considered as the suitable adjustment of coexisting soft and hard magnetic phases particularly in grain boundary regime of magnetic composites. Although origin of ferromagnetism in graphite based materials is yet to be understood, but most of the reports attributed grain boundary defects and breaking of covalent bonds in crystalline structure as the possible origins for the ferromagnetism of graphite [24-26]. Our work shows that not only magnetic particle concentration alone, but clay content (hardness) of the lead pencils and subsequent intercalated disorder in graphite structure [27, 28] plays a deterministic role in modified magnetism of the magnetic composite using lead pencils. As shown in Fig. 7, temperature dependence of ZFC and FC magnetization of 6BCF75 sample is drastically different from the 6B sample. For 6B sample at 500 Oe (higher field applied to avoid measurement error due to small magnetization) there is no significant separation between ZFC and FC curves, which has been reported in graphite samples [16]. The absence or very weak magnetic separation between ZFC and FC magnetization curves of our 6B sample is observed due to higher applied field than coercivity of the soft ferromagnetic sample [17, 29]. A low temperature increase of magnetization in 6B sample could be attributed to paramagnetic impurity or frustrated grain boundary spins [17]. On the other hand, magnetic measurement at 100 Oe (well below of the coercivity) of 6BCF75 sample exhibited a clear separation between ZFC and FC curves. The possibility of superparamagnetic blocking temperature above room temperature in composite 6BCF75 at 100 Oe is an interesting observation in comparison with 6B at 500 Oe, where there is no irreversibility. The blocking temperature of mechanical milled $CoFeO_4$ has been found well above room temperature [22]. As the temperature decreases below 330 K, ferromagnetic grains of $CoFeO_4$ were rapidly blocked in pencil matrix during zero field cooled mode, where as the field cooled curve is saturated below room temperature. This indicates that dipole-dipole interactions of the ferromagnetic grains are easily overcome by the field cooled mode. Viewing separation between ZFC and FC curves up to 330 K, we understand that superparamagnetic type blocking temperature of 6BFC75 magnetic composite is well above room temperature [22] and it could be tuned near to room temperature by adjusting applied magnetic field and content of lead pencils.

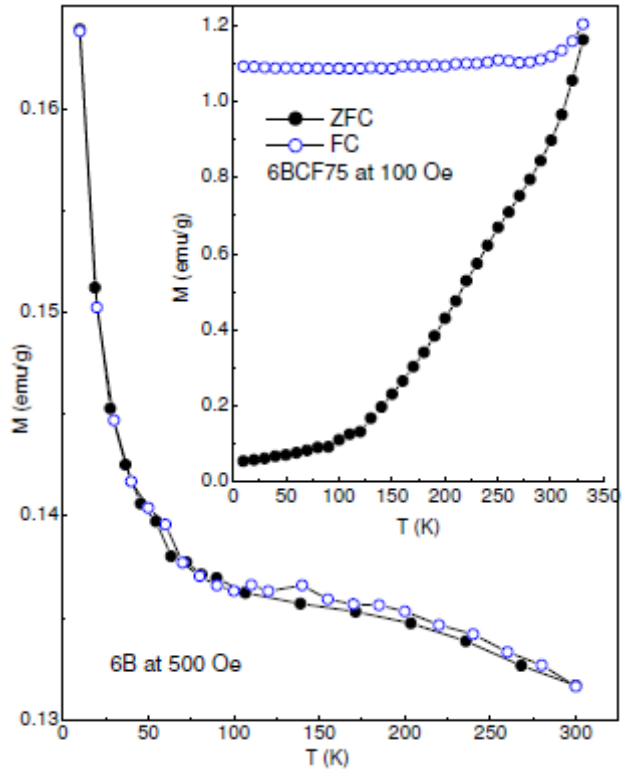

Fig. 7 (Colour online) Temperature dependence of ZFC (closed symbol) and FC (open symbol) magnetization of 6B (main frame) and 6BCF75 (inset) samples.

**4. Conclusions**

Lead graphite-pencils have been associated in our daily life but it could be an easy source for understanding many unexplained properties of graphite based materials. The present work shows that lead pencils belong not only to the class of heterogeneous materials, but also exhibited good soft ferromagnetic properties at room temperature. Its ferromagnetic parameters are largely controlled by grain boundary disorder and heterogeneity existing in the material. The presence of magnetic impurity Fe is acting as the catalyst for further enhancement of soft ferromagnetic parameters with increasing moment. Some of the lead pencils have been identified to be extremely useful component for making magnetic composite that provided wide scope for tuning ferromagnetic parameters. 6B pencil has been found to be the better candidate for achieving higher magnetization in the $CoFe_2O_4$ composite, where as 5H is the best candidate for controlling both magnetization and coercivity. The findings of this work are expected to be useful in understanding the mechanism of ferromagnetism in graphite based materials, especially to develop organic magnets for room temperature applications and tuning its magnetic parameters.

**Acknowledgments:** Author thanks to M.Sc. students Chiranjeev Das and B.S. Akila for helping in preparation and characterization of the samples. Author thanks to CIF, Pondicherry University, for providing experimental facilities. Author thanks to Asok Poddar for SQUID measurement of the samples. The financial support from UGC [F.NO. 33-5/2007 (SR)] is also acknowledged.